\documentclass[conference,compsoc]{IEEEtran}

\usepackage[utf8]{inputenc} 
\usepackage[T1]{fontenc}
\usepackage{url}
\usepackage{ifthen}
\usepackage{cite}
\usepackage[cmex10]{amsmath}

\usepackage{graphicx}
\usepackage{amsfonts}
\usepackage{amssymb}
\usepackage{mathrsfs}
\usepackage{mdwmath}
\usepackage{mdwtab}
\usepackage{dsfont}
\usepackage[ruled,vlined,linesnumbered]{algorithm2e} 
\DeclareMathOperator*{\argmax}{arg\,max}

\newcommand\smallO{
	\mathchoice
	{{\scriptstyle\mathcal{O}}}
	{{\scriptstyle\mathcal{O}}}
	{{\scriptscriptstyle\mathcal{O}}}
	{\scalebox{.7}{$\scriptscriptstyle\mathcal{O}$}}
}
\newtheorem{remark}{Remark}
\newtheorem{theorem}{Theorem}
\newtheorem{lemma}{Lemma}

\newtheorem{defi}{Definition}
\newtheorem{coro}{Corollary}
\newcommand{\tabincell}[2]{\begin{tabular}{@{}#1@{}}#2\end{tabular}}  

\begin{document}
\title{SENATE: A Permissionless Byzantine Consensus Protocol in Wireless Networks}

\author{\IEEEauthorblockN{Zhiyuan Jiang$^1$, Bhaskar Krishnamachari$^2$, Sheng Zhou$^1$,  Zhisheng Niu$^1$, \IEEEmembership{Fellow,~IEEE}}
    \IEEEauthorblockA{$^1$\{zhiyuan, sheng.zhou, niuzhs\}@tsinghua.edu.cn, Tsinghua University, Beijing, China\\
        $^2$ bkrishna@usc.edu, University of Southern California, Los Angeles, USA}}

    \maketitle

\begin{abstract}
The blockchain technology has achieved tremendous success in open (permissionless) decentralized consensus by employing proof-of-work (PoW) or its variants, whereby unauthorized nodes cannot gain disproportionate impact on consensus beyond their computational power. However, PoW-based systems incur a high delay and low throughput, making them ineffective in dealing with real-time applications. On the other hand, byzantine fault-tolerant (BFT) consensus algorithms with better delay and throughput performance have been employed in closed (permissioned) settings to avoid Sybil attacks. 
In this paper, we present Sybil-proof wirelEss Network coordinAte based byzanTine consEnsus (SENATE), which is based on the conventional BFT consensus framework yet works in open systems of wireless devices where faulty nodes may launch Sybil attacks. As in a Senate in the legislature where the quota of senators per state (district) is a constant irrespective with the population of the state, ``senators'' in SENATE are selected from participating distributed nodes based on their wireless network coordinates (WNC) with a fixed number of nodes per district in the WNC space. Elected senators then participate in the subsequent consensus reaching process and broadcast the result. Thereby, SENATE is proof against Sybil attacks since pseudonyms of a faulty node are likely to be adjacent in the WNC space and hence fail to be elected.
\end{abstract}

\section{Introduction}
\label{sec_intro}
Recent years have witnessed an explosive development in digital cryptocurrency, both in academic fields and financial markets. Behind its tremendous success, the key enabling technology of digital cryptocurrency is the blockchain \cite{bitcoin,tch16} which combines several judiciously designed techniques to facilitate trusted distributed ledgers such that intermediary can be eliminated during transactions. In particular, the Bitcoin blockchain ingeniously adopts the proof-of-work (PoW) for mining to, among other purposes, deal with identity attacks (Sybil attacks) in open (permissionless) systems wherein the identities of participating nodes are not assumed to be known a priori. Specifically, each node, whether it is authentic or a pseudonym, must solve a cryptographic puzzle to participate in the block generation process and obtain rewards (mining). Therefore, the impact of a mining node is directly tied to its computational power, irrespective of the number of identities it has. 

\begin{figure}[!t]
	\centering
	\includegraphics[width=0.5\textwidth]{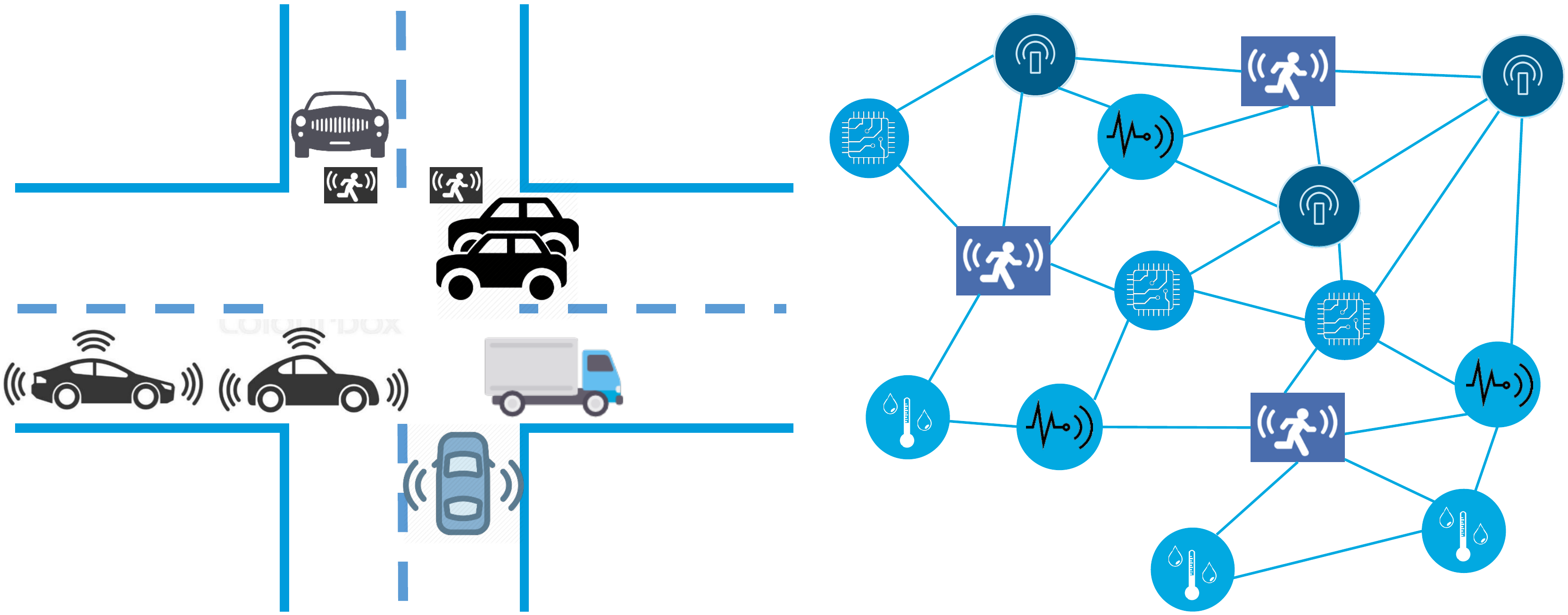}
	\caption{Time-critical application scenarios, e.g., autonomous driving systems wherein vehicles and pedestrians go through intersections based on distributed consensus (left); Internet-of-Things where terminals (drones, sensors and actuators) act based on coordinated and synchronized behavior (right).}
	\label{Fig_app}
\end{figure}

According to the necessity of a prior identity authorization procedure, blockchain technologies can be categorized by permissionless and permissioned blockchains. Permissionless blockchains, such as Bitcoin \cite{bitcoin} and Ethereum \cite{iddo14}, are applied in open systems wherein faulty nodes may apply Sybil attacks, the counteraction of which usually involves enforcing a strict coupling between the consensus impact of a node and the computational power (PoW for Bitcoin) or the resources (proof-of-stake for Ethereum Casper) of the node. Despite the robustness against Sybil attacks, a price is payed that permissionless blockchains usually suffers from high processing delay ($6$ blocks are recommended in Bitcoin which amounts to one hour) and low throughput (at most $7$ transactions per second \cite{Gil17} for Bitcoin). Many existing works try to remedy this issue \cite{Gil17,eyal16}, however, the inherent mining based probabilistic consensus reaching technique is the key limiting factor.
On the other hand, permissioned blockchains, such as Hyperledger Fabric \cite{cac16}, need not be wary of Sybil attacks since all participating nodes have gone through an explicit authentication process such that they can be trusted. Adopting a long line of existing research on byzantine fault-tolerant (BFT) protocols \cite{pbft,Diego14,Gue10}, the processing delay and throughput of permissioned blockchains can be dramatically improved; a nice comparison between traditional BFT protocols and permissionless blockchains is presented by Vukoli{\'{c}} \cite{vuk16}. 

\begin{table}[!t]
	\caption{Comparisons with blockchain technologies}
	\label{tab_byz}
	\centering
	\begin{tabular}{| c | c | c | c |}
		\hline
		 & \tabincell{c}{Permissionless \\blockchain} & \tabincell{c}{Permissioned\\ blockchain} & SENATE  \\
		\hline
		Open system & Yes & No & Yes \\
		\hline
		Delay & Large & Small & Small \\
		\hline
	\end{tabular}
\end{table}

In this paper, a large-scale dense wireless network scenario is considered, which is likely to be encountered in the future in Internet-of-Things deployments as well as in the context of vehicle networks and intelligent transportation (Figure \ref{Fig_app}). We come up with a solution for Sybil-proof BFT consensus which offers the benefits of both permissioned and permissionless blockchains (cf. Table \ref{tab_byz} for details), i.e., low-delay and high-throughput BFT consensus in permissionless systems. The proposed protocol, namely Sybil-proof wireless network coordinate (WNC) based byzantine consensus (SENATE), consists of three major phases: sortition, senator selection and byzantine agreement. SENATE thwarts the Sybil attack by exploiting the fact that even a faulty node cannot forge its wireless channel to other nodes such that a unique wireless fingerprint can be leveraged to identify nodes; a fully decentralized approach is proposed to achieve this.

\textbf{Notations}: Throughout the paper, we use boldface uppercase letters, boldface lowercase letters and lowercase letters to designate matrices, column vectors and scalars, respectively. The transpose of a matrix is denoted by $(\cdot)^{\mathsf{T}}$. $X_{i,j}$ and $x_i$ denotes the $(i,j)$-th entry and $i$-th element of matrix $\boldsymbol{X}$ and vector $\boldsymbol{x}$, respectively. The $\ell_p$ and nuclear norm of a matrix are denoted by $\|\cdot\|_p$ and $\|\cdot\|_*$, respectively. The vector consisting of the diagonal entries of a square matrix is denoted by $\mathsf{diag}[\cdot]$. The trace of a matrix is denoted by $\mathsf{tr}[\cdot]$. The matrix with all entries being one is denoted by $\boldsymbol{1}$, and likewise zero matrix is denoted by $\boldsymbol{0}$.

\subsection{Related Work}
The idea of only allowing selected nodes to participate in BFT consensus reaching is shared by, e.g., NEO \cite{neo} and Algorand \cite{Gil17}. NEO is a delegated BFT consensus based blockchain in which a small number of servers are \emph{statically} configured to run consensus on behalf of a larger open network. Similar with SENATE, Algorand counteracts Sybil attacks by adopting a sortition phase; the difference is that a random verifiable function based solution combining with proof-of-stake (PoS) is leveraged by Algorand, whereas SENATE is based on the underlying wireless channels. Because SENATE is not employing PoS, it is not tied to a digital currency and thus can be applied more broadly to achieving consensus in wireless systems; moreover, it avoids the unfairness introduced by PoS which intentionally favors participants with more resources.

Proof-of-location (PoL) in wireless networks is adopted by Dasu \emph{et al.} \cite{tam18} to replace PoW for faster transactions. The PoLs are generated by authorities such as wireless network operators and hence some notion of centralization is introduced. SENATE also uses the concept of PoL whereas, on a high level, nodes generate PoLs in a fully-decentralized manner without any trusted authority.

There have been some work where wireless channel fingerprints are utilized to protect against Sybil attacks (cf. \cite{new04} for a survey and \cite{far06,Dem06} for a signalprint-based approach). However, most existing work relies on a trusted authority to verify the wireless channels of nodes \cite{far06}, or pre-distribute encrypted keys \cite{new04}. In \cite{Dem06}, a wireless ad hoc network of commodity $802.11$ devices is considered; a view selection policy based on signalprint observations is proposed. Our work considers a fully decentralized wireless network and a novel WNC based protocol (SENATE) is proposed; compared with existing work, SENATE has much lower running time and better understandability and hence more favourable for real-world implementations.

\section{System Model}
\label{sec_model}
We consider a wireless network consisting of $N$ geographically distributed nodes with full connectivity, namely any pair of nodes in the network are within each other's radio communication range. The system is open, or permissionless, in the sense that any node can join the system without prior identity authentication. The objective is for the \emph{good} nodes to reach a valid (the definition for validity is addressed later) consensus on a set of values over a certain time period such that deterministic concerted actions can be carried out, in the meantime, subject to malicious behaviors by \emph{faulty} nodes. Note that the considered scenario distinguishes from the state machine replication wherein a log is proposed by a client and different nodes agree on the same log record; here different nodes may have different set of initial values, e.g., sensory data from environment, and hence a reasonably good (valid) consensus needs to be reached.

Unlike existing work on byzantine  consensus which mainly adopts the Internet as the overlay network, a wireless overlay is considered. In this regard, the behavior of a faulty node should be clarified. Specifically, the following assumptions are made in this paper.
\begin{itemize}
    \item
    The objective of a faulty node is to \emph{rig} the consensus reaching process to benefit itself, rather than halting the process. 
    \item
    To achieve its purpose, possible malicious behaviors include: $(1)$ Byzantine node \cite{pbft}, namely it does not comply with the protocol and can report arbitrary messages; $(2)$ Sybil attack \cite{sybil}: it can generate pseudonyms to gain inappropriate power in the process of reaching consensus. 
    \item
    In the overlay wireless network, a faulty node does not block or interfere with other nodes' transmissions and messages.    
\end{itemize}

The first assumption describes the motive of a faulty node and therefore has implications on the other two assumptions. The second assumption simply states that, on a message level, there is no restriction on the behavior of a faulty node, both from the perspectives of the message content and the identity of the message sender. In most existing byzantine agreement protocols with the Internet as overlay \cite{pbft}, the third assumption is also implied which limits the malicious behavior of a node to itself; whereas in wireless networks with the broadcast nature of electromagnetic waves, this assumption has more implications, meaning that a faulty node is assumed to comply with the communication protocol. For instance, a faulty node would not transmit when another node is scheduled (by the consensus protocol). This assumption stems in large part from the first assumption, since messing with the communication protocol, e.g., transmitting with a high power and thus blocking other nodes, leads to retransmissions and hence halting the consensus reaching process. Besides, the following two reasons also justify the assumption: $(1)$ an attack becomes trivially devastating without this assumption, namely a faulty node with sufficient transmit power can block other transmissions all the time to prevent reaching consensus; $(2)$ a node not complying with the communication protocol is obviously malicious and easy to spot.

In this work, we assume nodes can obtain ranging estimations based on others' pilot signals. However, we do not focus on specific methods to obtain the distance estimations; they can be based on receive signal strength (RSS), time of arrival (ToA) or other approaches which have been studied extensively \cite{pat05}. The net effect of ranging estimations is considered by a statistical model, i.e., $\hat d_{ij} = \sigma_{ij} d_{ij} + n_{ij}$, where $d_{ij}$ denotes the geographic distance between node-$i$ and node-$j$ and hence $d_{ji}=d_{ij}$, the distance estimation at node-$j$ from node-$i$ is denoted by $\hat d_{ij}$, the estimation error is introduced by multiplicative and additive random coefficients $\sigma_{ij}$ and $n_{ij}$, respectively. In the wireless localization literature, it is usually assumed that
\begin{itemize}
    \item 
    {For ToA-based ranging estimations, $\sigma_{ij}=0$ and $n_{ij}$ is modeled as a Gaussian distributed variable.}
    \item
    For RSS-based ranging estimations, the shadow fading is modeled as $\sigma_{ij}$ which is assumed to be Gaussian distributed; it is often termed as log-normal shadow fading by taking logarithm on both sides.    
\end{itemize}
Considering the ranging estimation error, it is observed thereby that the RSS-based approach is effective with short distances since there is a multiplicative error component; the ToA-based approach applies to a more wide range of distances although it may require a central node to calibrate the clocks of terminals to ensure synchronization.

\section{Overview}
\label{sec_ov}
SENATE consists of three major phases: sortition, senator selection and byzantine agreement. 

\subsection{Sortition}
In the sortition process, the objective is to prevent faulty nodes to generate \emph{arbitrarily} many pseudonyms; note that this does not mean we eliminate the Sybil attack by sortition completely. The key to achieve this is by developing an ALOHA game with selfish users \cite{mac01} such that no one can cheat based on the Nash equilibrium arguments. 

In the ALOHA game, a critical requirement is that every good player (node) knows the total number of users (including faulty nodes) in the system to determine its action; in the considered scenario, this requirement poses a challenge in the presence of faulty nodes, namely faulty nodes can let good nodes believe there are fewer nodes in the system such that good nodes may behave more conservatively and hence benefit faulty nodes. To prevent this, a \emph{chorus} procedure is proposed which leverages the unique pattern of channel power delay profile (PDP) to estimate the total number of participating nodes. The key observation is that a faulty node cannot forge multi-path components (MPCs) and hence this feature can be utilized, especially with line-of-sight (LoS) transmission environment, to estimate the node quantity.

\textbf{Chorus}: In this procedure consisting of $T$ time slots, each node is supposed to randomly select one time slot to receive and, on the other hand, transmit pilot symbols in the remaining $T-1$ time slots. An analysis in Section \ref{sec_sort} shows that even in the presence of faulty nodes, this procedure is robust. In the receive time slot of, e.g., node-$i$, it estimates the PDP of receive signal and calculates the number of MPCs. In an LoS environment, this measurement gives an accurate estimation of the number of transmitters in the system; moreover, even a faulty node cannot generate multiple MPCs given its location. 

This is analogous to let all nodes perform chorus first such that every node can have an estimation of the population based on nodes' unique timbre. The procedure lasts $T$ time slots since we assume that nodes cannot operate in the full-duplex mode; otherwise the procedure can be shortened to one time slot wherein every node transmit and receive simultaneously. The detailed procedure, as well as the analysis for appropriate $T$, is given in Section \ref{sec_sort}.

\textbf{ALOHA game based sortition}: Given each node's estimations about the total number of nodes in the system, we let every node be selfish in the ALOHA game to prevent faulty nodes from gaining advantages; this all-be-selfish methodology is essentially identical with the blockchain technology which allows all miners to compete for the opportunity to register a block. In particular, an ALOHA random access game with selfish users is implemented. It is roughly described as follows.

\begin{itemize}
    \item 
    Every node is selfish, in the sense that they all want to transmit as soon as possible in a collision-free time slot. However, after a successful transmission, a good node would stop competing whereas a faulty node might keep on transmitting to launch Sybil attacks.
    \item 
    Once a node successfully transmits in a time slot (collision-free), it is selected as the $s$-th candidate where $s \in \{1,...,S\}$ denotes the $s$-th successful transmission in the ALOHA game; then it transmits a pilot signal for ranging estimations immediately afterwards.
\end{itemize}

By this definition, it can be shown (Section \ref{sec_sort}) that a Nash equilibrium exists based on which every node adopts the same mixed-strategy \cite{mac01}: transmit in each time slot with a probability $p$ with $0<p<1$, and no one can benefit by changing the strategy unilaterally. 

After a quorum of $S$ candidates is reached, the sortition phase terminates with $S$ candidates going into the senator selection in the next phase. Note that a Sybil attack is still possible; a faulty node may occupy several seats among the candidates.

\subsection{Senator Selection}
This phase is dedicated to further removing the pseudonyms generated by faulty nodes, by cross-checking the ranging estimations among nodes in a fully distributed manner.

After $S$ candidates are selected, they no longer follow the ALOHA-based random access protocol. Instead, each candidate is assigned a unique time slot to transmit in a frame of $S$ time slots in this phase.

Since every candidate has transmitted a pilot signal, every candidate has obtained the distance estimations from other candidates. For candidate-$i$, its distance estimations are denoted by a vector
\begin{equation}
    \hat{\boldsymbol{d}}_i = \left[\hat{d}_{1i},...,\hat{d}_{(i-1)i},0,\hat{d}_{(i+1)i},...,\hat{d}_{Si}\right]^\mathsf{T},\,i=1,...,S.
\end{equation}
The estimated Euclidean distance matrix (EDM) \cite{dok15} with squared norm is hence
\begin{equation}
    \hat{\boldsymbol{D}} \triangleq \left[\hat{\boldsymbol{d}}_1^2,...,\hat{\boldsymbol{d}}_S^2\right].
\end{equation}

\textbf{Distance feedback and symmetry verification}: Each candidate feeds back its $\hat{\boldsymbol{d}}_i$ in its dedicated time slot. Afterwards, every candidate obtains the distance estimations between any pair of candidates (double-directional) in the network. Note that $d_{ij} = d_{ij}$, $\forall i,j$, and thereby every node can remove suspicious distance feedback based on checking the estimated EDM (here we assume the feedback is perfect)
\begin{equation}
\label{reciprocity}
    \left|\hat{d}_{ij}^2 - \hat{d}_{ji}^2\right| < \epsilon
\end{equation}
where $\epsilon$ is a constant related to $\sigma_{ij}$ and $n_{ij}$. In this case, as long as \eqref{reciprocity} does not hold, both $\hat{d}_{ij}$ and $\hat{d}_{ij}$ are removed since we cannot tell if node-$i$ or node-$j$ is lying. 

\textbf{Robust WNC generation}: Despite the fact that the symmetry verification can, to some extent depending on distance estimation error, eliminate untruthful distance feedback, a faulty node can still launch what we refer to as a ``shout attack''\footnote{Likewise, a ``whisper attack'' can be defined by which the faulty node pretends to be nearer to other nodes. For ease of exposition, we use the shout attack for illustration henceforth.}.
\begin{defi}[Shout Attack]
A shout attack is that a faulty node pretends to be further away to other nodes, by synchronously adding to the distance estimations to other nodes. In particular, for ToA-based ranging estimations, a faulty node can purposely transmit pilot signals later than supposed, and, accordingly, feed back tampered (larger) distance estimations; for RSS-based ranging estimations, a faulty node can purposely amplify its pilot signal power and, accordingly, feed back tampered (larger) distance estimations. $\hfill\square$
\end{defi}

By definition, a shout attack cannot be detected by symmetry verification and gives a faulty node arbitrarily many fake geographical locations that are arbitrarily far from its real one. The purpose of a shout attack is hence to create pseudonyms and facilitate the Sybil attack, which causes a severe challenge to SENATE since SENATE uses the location information for Sybil protection. 

\begin{figure}[!h]
	\centering
	\includegraphics[width=0.42\textwidth]{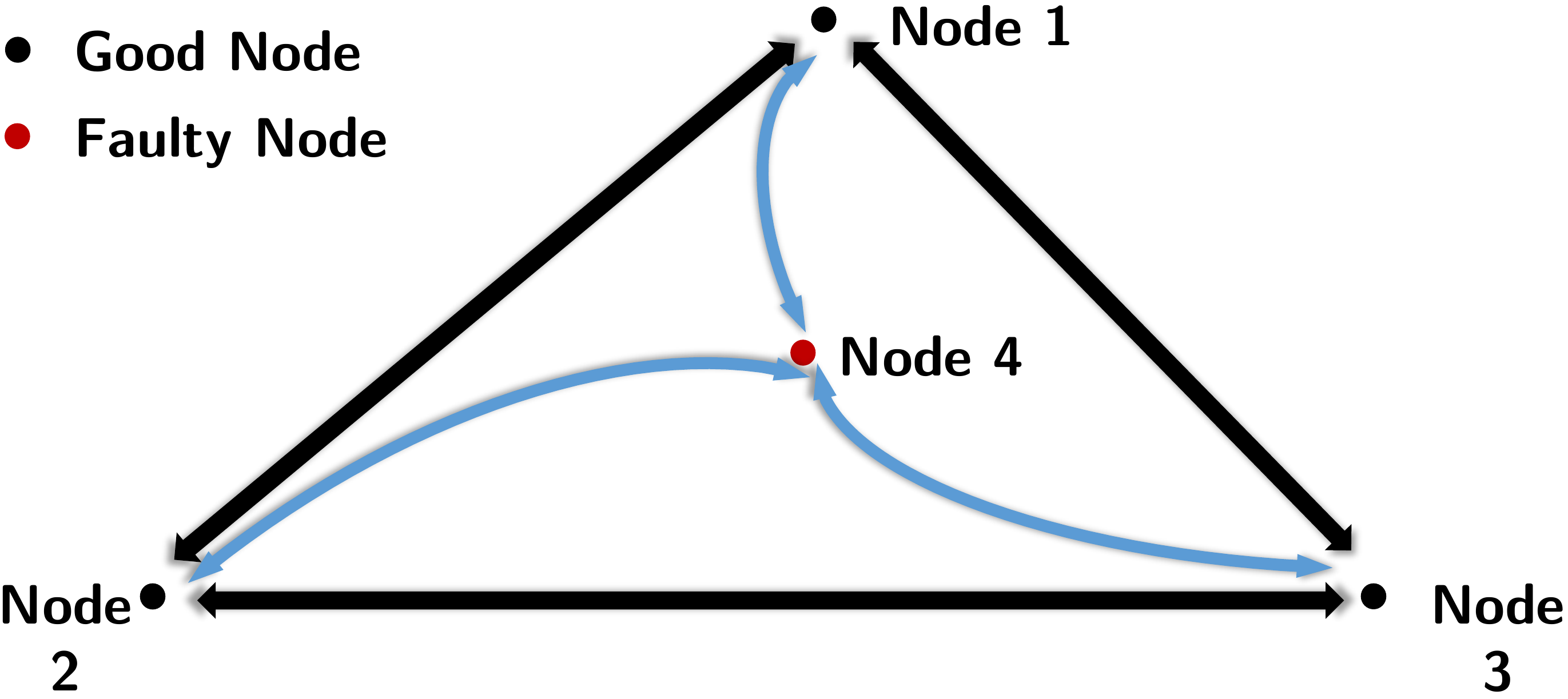}
	\caption{A seesaw test. A faulty node launching the shout attack would be ``levered out''.}
	\label{Fig_seesaw}
\end{figure}

To thwart the shout attack, we introduce the \textbf{seesaw test} based on the following intuition. In the real world with (at most) $3$-dimensional space, it is increasingly unlikely that a faulty node, which launches the shout attack, is further away to other nodes proportionally, as the number of nodes grows. This is analogous to placing elastic sticks between each pair of nodes in the system, with the lengths of sticks given by the distance estimations. The circumstance for a faulty node launching shout attack in the $2$-dimensional space is illustrated in Figure \ref{Fig_seesaw}; its related sticks are bent dramatically and hence the elastic force levers it out (screened out by the seesaw test), like being on the lighter side of seesaws. This argument is mathematically formalized in Theorem \ref{thm_seesaw} which states the local error is proportional to the number of good nodes.

An iterative WNC calculation with seesaw tests is proposed, the essence of which can be illustrated as follows. In each round of the iteration, every node, e.g., node-$i$, uses the distance estimation to another node-$j$ to push (resp. pull) node-$j$ if the distance estimation is larger (resp. smaller) than the predicted distance between the two nodes by the current WNC; then node-$j$ moves accordingly. At the end of each round, the node with the largest local error, i.e., its related seesaws are bent the most, is identified as a faulty node and therefore removed; note that a termination criterion (specified later) is added such that the WNC calculation terminates when the system has small prediction error. 

\textbf{$K$-means clustering}. After obtaining the coordinates of all candidates, a $K$-means clustering algorithm \cite{mac67} is applied to the coordinates such that all candidates are divided into $K$ clusters. Then $K$ representatives, each from one cluster, are selected as senators; this prevents Sybil attacks because pseudonyms of one faulty candidate are likely to fall into the same cluster. 

\subsection{Byzantine Agreement}
The $K$ senators run a byzantine agreement protocol to reach consensus. We primarily consider the median validity for consensus, which is defined as follows \cite{sto16}.\footnote{Nevertheless, basically any BFT protocol can be plugged in to SENATE in this phase, and works well in an open system since we have achieved Sybil-proof in the previous phases.} Assume a single consensus value is to be reached upon. Denote by $G$ the sorted array of the initial values of good nodes. Among $n$ nodes, $f$ nodes are faulty and it is assumed that $f \le t$ and hence $G = [G[0],...,G[n-f-1]]$. 
\begin{defi}[Median Validity]
We call a value $x$ median-valid, if it holds that 
\begin{equation}
    G\left[\left\lceil\frac{n-f}{2}\right\rceil -1 -t\right] \le x \le G\left[\left\lceil\frac{n-f}{2}\right\rceil -1 +t\right]. \hfill\square
\end{equation}
\end{defi}
Thereby, we adopt the Jack algorithm \cite{sto16} which ensures the following properties, as long as the number of faulty senators, i.e., $f$, satisfies $K \ge 3f+1$.
\begin{itemize}
    \item 
    \textbf{Agreement}: For every selection of input values and every selection of faulty senators, all good senators can decide on the same value.
    \item 
    \textbf{Termination}: Every good senator can decide on a value in finite time.
    \item 
    \textbf{Median Validity}: The decision is median-valid.
\end{itemize}

Upon agreement, every senator broadcasts its consensus value, and every good node in the system adopts the majority value. Since consensus is reached in the senate, the majority value can reflect the consensus value and ensures safety and agreement among all nodes.

\section{Sortition}
\label{sec_sort}
The sortition phase is described in Algorithm \ref{alg:sort}. We elaborate on several details as follows.
\begin{algorithm}[!t]
	\caption{Sortition}
	\label{alg:sort}
		\KwIn{Node-$1,...,N$;}
		\KwOut{Candidate-$1,...,S$;}
		\textbf{{Chorus Procedure}}\\
		Every node (node-$i$) uniform-randomly selects $1 \le t_i \le T$.\\
		\For{t=1:T}{
		\If{$t=t_i$}{Node-$i$ transmits a pilot symbol in the $t$-th time slot.}
		\Else{Node-$i$ receives signals and estimates the number of transmitting nodes based on the receive PDP; the estimation is denoted by $\sigma_i$.\\
		The total number of nodes estimated by node-$i$ is $\hat N_i = 1 + \frac{T}{T-1}\sigma_i$.
		}}
		\textbf{ALOHA Game based Sortition}\\
		Every node can transmit in every time slot.\\
		\For{s=1:S}{The $s$-th candidate is selected when a new successful transmission happens and the corresponding node is candidate-$s$; afterwards, it transmits a pilot signal.}
		\Return Candidate-$1,...,S$.
\end{algorithm}

\subsection{Analysis on Chorus Duration}
When a node, e.g., node-$i$, receives signals, assuming its estimation on the number of transmitting nodes is correct, the probability of a good node transmitting in this time slot is 
\begin{equation}
    p_j = 1 - \frac{1}{T},\,\textrm{node-$j$ is a good node and $j \neq i$.}
\end{equation}
Therefore, the unbiased estimate should be
\begin{equation}
    \hat N_i = 1 + \frac{T}{T-1} q_i,
\end{equation}
where $q_i$ denotes the estimation of transmitting node in node-$i$'s receive time slot. The optimal attack a faulty node can launch is to let good nodes believe there are more nodes in the system such that the transmission strategy of the latter would be more conservative. Therefore, the worst effect all faulty nodes can conjure is by always transmitting, and thereby,
\begin{equation}
    \hat N_i \le 1+ \frac{T}{T-1} F + \frac{T}{T-1} \sum_{n=1}^{N-F-1}m_n,
\end{equation}
where $m_n$ is a Bernoulli random variable with parameter $1 - \frac{1}{T}$. By letting $T$ be sufficiently large compared with $N$, then
\begin{IEEEeqnarray}{rCl}
    \hat N_i &\le& 1+ \frac{T}{T-1} F + (N-F-1) + \smallO\left(\frac{N}{T}\right) \nonumber\\
    &=& N + \frac{F}{T-1} + \smallO\left(\frac{N}{T}\right),\,F \le N
\end{IEEEeqnarray}
where $\smallO\left(\cdot\right)$ denotes infinitesimal. In an LTE system with time slot of $0.5$~{ms} and $100$ nodes, a chorus procedure lasting $1$~s yields $\frac{N}{T}=0.05$.
\subsection{Nash Equilibrium of ALOHA Game}
An ALOHA game is described as follows. Every node participates in the game.
\begin{itemize}
    \item In a time slot a node successfully transmits (collision-free), the node receives a payoff of $1-c$ where $c \in [0,1]$ denotes the one-time transmission cost and leaves the game.\footnote{A faulty node may otherwise stay in the game and keep playing.}
    \item In a time slot if a collision happens, every transmitting node receives a payoff of $-c$.
\end{itemize}

A detailed payoff function is described in Table \ref{tab_game}. Every node's goal is to maximize its payoff in a single time slot and the game is repeated. Based on the game setting, we can prove the existence of Nash equilibrium.
\begin{theorem}
\label{thm_nash}
There exists a Nash equilibrium that every node adopts the same mixed-strategy: transmit with probability $p(c,N)$ in each time slot, where 
\begin{equation}
\label{p_n}
    p(c,N) = 1-\sqrt[N-1]{c}.
\end{equation}
\end{theorem}
\begin{IEEEproof}
The proof is based on \cite{mac01}. See Appendix \ref{app_nash} for details.
\end{IEEEproof}
\begin{remark}
Theorem \ref{thm_nash} indicates that by allowing every node to be selfish, a symmetric equilibria exists, namely even a faulty node cannot improve its payoff by changing its transmission probability unilaterally. Specifically, if it increases its transmission probability, there would be more collisions and the payoff decreases due to cost $c$; if it decreases its chance then its success chance also decreases. The transmission cost per time slot $c$ clearly plays a critical role here, which denotes the relative cost per transmission as compared with one successful transmission. In practice, we propose that, aside from the power and resource cost due to wireless transmissions, an economic approach can be applied whereby a small fee is charged for every sortition transmission to enhance the robustness of the process.  
\end{remark}
\section{Senator Selection}
\label{sec_senator}
\begin{algorithm}[!h]
	\caption{Senator Selection}
	\label{alg:wnc}
		\KwIn{Candidate-$1,...,S$;}
		\KwOut{Senator-$1$,...,$K$; $\mathsf{validSenate}$.}
		\textbf{{Distance estimation and feedback}}\\
		\For{i=1:S}{Based on the pilot signals received in the sortition phase, candidate-$i$ estimtes its distance to other candidates and feeds back its distance vector $\hat{\boldsymbol{d}}_i$.}
		\textbf{{EDM symmetry verification}}\\
		$\boldsymbol{\Sigma} \triangleq |\hat{\boldsymbol{D}}-\hat{\boldsymbol{D}}^\mathsf{T}|$.\\
		\For{Every element $D_{ij}$ in $\hat{\boldsymbol{D}}$}{
		\If{$\Sigma_{ij} > \Delta d$}{$\hat{D}_{ij} \leftarrow \mathsf{invalidValue}$.}
		}
		\textbf{{Robust WNC generation}}\\
		At every terminal, generate the WNC simultaneously based on the same following procedure:\\
		$\mathsf{terminate} \leftarrow \mathsf{false}$; $\boldsymbol{X} \triangleq [\boldsymbol{x}_1,...,\boldsymbol{x}_S]^\mathsf{T} \leftarrow \boldsymbol{0}_{S \times 2}$; $\boldsymbol{e} \leftarrow \boldsymbol{0}_{S \time 1}$.\\
		\While{$\mathsf{terminate} = \mathsf{false}$}{
		\For{$\{i,j\} \in \{1,...,S\} \times \{1,...,S\}$}{
		\If{$D_{ij} \neq \mathsf{invalidValue}$ or $0$}{
		
		$w \leftarrow \frac{e_i}{e_i + e_j}$; $e_i \leftarrow \frac{|\|\boldsymbol{x}_i-\boldsymbol{x}_j\|_2 - \hat{d}_{ij}|}{\hat{d}_{ij}} \delta w + (1-\delta w) e_i$; $\boldsymbol{f} \leftarrow  w \left(\frac{|\|\boldsymbol{x}_i-\boldsymbol{x}_j\|_2 - \hat{d}_{ij}|}{\hat{d}_{ij}}\right) (\boldsymbol{x}_i-\boldsymbol{x}_j)$; \\ 
		\label{viv}
		$\boldsymbol{x}_i \leftarrow \boldsymbol{x}_i + \gamma \boldsymbol{f}$.
		}}
		$\mathsf{maxError} \leftarrow \max_i\{e_i\}$; $\mathsf{errIndex} \leftarrow \argmax_i\{e_i\}$.\\
		\If{$\mathsf{maxError} > \beta \frac{\sum{e_i}}{S}$}{
		Remove candidate-$\mathsf{errIndex}$, and its corresponding entries in $\boldsymbol{X}$, $\boldsymbol{e}$ and $\hat{\boldsymbol{D}}$; $S \leftarrow S-1$.
		}
		\Else{$\mathsf{terminate} \leftarrow \mathsf{true}$}
		}
		\textbf{{$K$-means clustering}}\\
		\If{$S<K$}{$\mathsf{validSenate} \leftarrow \mathsf{false}$.}
		\Else{$\boldsymbol{S} \leftarrow \mathsf{Kmeans} (\boldsymbol{X},K)$; $\mathsf{validSenate} \leftarrow \mathsf{true}$
		}
		\Return $\boldsymbol{S}$; $\mathsf{validSenate}$.
\end{algorithm}
In this phase, the $S$ candidates transmit in a round-robin fashion. The detailed algorithm description of the phase is presented in Algorithm \ref{alg:wnc}, and some explanations follow.
\subsection{Robust WNC Generation}
The rationale for robust WNC generation is as follows. In the face of EDM estimation error introduced by faulty nodes, i.e., denote
\begin{equation}
\label{10}
    \hat{\boldsymbol{D}} = \boldsymbol{D} + \boldsymbol{E},
\end{equation}
where the entries of $\boldsymbol{E}$ can be arbitrarily large considering malicious behavior, our goal is to recover $\boldsymbol{D}$. Towards this end, two structures can be exploited: (a) although the faulty nodes can cause arbitrarily large error, the error is sparse in terms of entries of $\boldsymbol{E}$, i.e., majority is still good; (b) the EDM stems from space of limited dimensionality\footnote{We use $2$-dimensional space for ease of exposition in this paper. However, the generalization to $3$-dimensional is considered straightforward.} and hence there are mature tools in distance geometry \cite{dok15} that can be utilized to verify its authenticity. Thereby, considering $2$-dimensional space, the EDM can be written as
\begin{equation}
\label{9}
    \boldsymbol{D} = -2\boldsymbol{X}\boldsymbol{X}^\mathsf{T} + \boldsymbol{1} \mathsf{diag} \left(\boldsymbol{X}\boldsymbol{X}^\mathsf{T}\right)^\mathsf{T} + \mathsf{diag} \left(\boldsymbol{X}\boldsymbol{X}^\mathsf{T}\right) \boldsymbol{1}^\mathsf{T},
\end{equation}
where $\boldsymbol{X} \in \mathbb{R}^{S \times 2}$ is the geographical location coordinates of candidates, i.e., $\boldsymbol{X} \triangleq \left[\boldsymbol{x}_1,...,\boldsymbol{x}_S\right]^\mathsf{T}$. We then formulate the WNC generation problem as follows, exploiting the sparse error property.
\begin{flalign}
\label{p1}
\textbf{P1:}&&\mathop{\textrm{minimize}}\limits_{\boldsymbol{X},\boldsymbol{E}}  \|\boldsymbol{E}\|_0,\,\textrm{subj. } \, \eqref{10},\,\eqref{9} \textrm{ and }\mathsf{rank}(\boldsymbol{X})=2.&&
\end{flalign}
The ${\ell}^0$ norm based formulation in \textbf{P1} is notoriously non-convex and in fact NP-complete based on compressive sensing theory. Therefore, the ${\ell}^0$ norm in \textbf{P1} is relaxed to ${\ell}^1$ norm which often exhibits near optimal performance \cite{can06}, i.e.,
\begin{flalign}
\label{p2}
\textbf{P2:}&&\mathop{\textrm{minimize}}\limits_{\boldsymbol{X},\boldsymbol{E}}  \|\boldsymbol{E}\|_1,\,\textrm{subj. } \, \eqref{10},\,\eqref{9} \textrm{ and }\mathsf{rank}(\boldsymbol{X})=2.&&
\end{flalign}
We adopt a data-driven gradient-descend-based method to solve \textbf{P2}. Based on an estimation $\hat{d}_{ij}$, we can update $\boldsymbol{x}_i$ (or $\boldsymbol{x}_j$) based on the gradient of the objective function in \textbf{P2}:
\begin{equation}
    \boldsymbol{x}_i \leftarrow \boldsymbol{x}_i + \mu \frac{\partial\left|\|\boldsymbol{x}_i-\boldsymbol{x}_j\|_2^2-\hat{d}_{ij}^2\right|}{\partial \boldsymbol{x}_i},
\end{equation}
which corresponds to the \ref{viv}-step in Algorithm \ref{alg:wnc}. Also note that in the algorithm, we keep track of the local error array $\boldsymbol{e}$ whose element $e_i$ represents the squared distance error related to candidate-$i$; that is how much candidate-$i$ is levered in the seesaw test (Figure \ref{Fig_seesaw}). Therefore in the \ref{viv}-step, we take into account the fact that a candidate with small error should not be updated based on the location of a candidate with large error; the latter is likely to be a faulty node. Based on this argument, we remove the candidate with the largest error at the end of each round, until the error is evenly distributed among candidates which means the error is introduced by ranging estimation instead of faulty nodes. 

In the case that the selected senators do not reach a quorum of $K$, Algorithm \ref{alg:wnc} returns $\mathsf{validSenate=false}$.

Intriguingly, this method is similar with the spring network based method where any pair of nodes are connected by a spring in, e.g., \cite{pat05,vivaldi}; the objective in those works is to minimize the elastic potential energy of the system (equivalent with the total square error (TSE) of distance prediction) given the current lengths of springs (distance estimations) by placing the nodes (the distances among nodes are the rest lengths of the springs) on a plane. Although the objective in \textbf{P2} is not minimizing the TSE, the presented data-driven gradient-descend-based method turns out to be similar with the Vivaldi algorithm \cite{vivaldi}, except for the faulty detection.
\subsection{Analysis on Seesaw Test}
The seesaw test is based on the rationale that a faulty node implementing the shout attack can be detected because its resultant location would be out of the $2$D space. A question arises accordingly: how \emph{out-of-space} the faulty node is given a certain strength of its shout attack, and moreover the effect of the number of good nodes. This question is important because its answer can quantitatively characterize the effectiveness of the seesaw test against forged locations.

In seeking for a concise and illustrative answer, we consider a simplified scenario where there is one faulty node, without loss of generality located at $\boldsymbol{x}_0 = (0,0)$, who is trying to launch a shout attack to $M$ good nodes located at $\boldsymbol{x}_m = (x_m,y_m)$, $\forall m \in {1,...,M}$. Concretely, we consider that the faulty node adds an arbitrary (independent with real node-locations) error vector to the entries in the EDM that are related to it; note that this is more general than the shout attack whereby the error is added synchronously. The arbitrary error is written based on \eqref{10} as
\begin{equation}
\label{e}
    \boldsymbol{E} = \left[ {\begin{array}{*{20}{c}}
0 & {{\boldsymbol{e}^\mathsf{T}} }\\
{\boldsymbol{e}} & \boldsymbol{0}_{M \times M}
\end{array}} \right].
\end{equation}
Note that no ranging estimation error is considered in this subsection to focus on the synthetic error by the faulty node. The level of out-of-space of the faulty node is measured by 
\begin{equation}
\label{test}
    h (\varsigma^2 ) \triangleq \mathbb{E}_{\boldsymbol{X}} \left[ \min_{\boldsymbol{Z},\,\mathsf{rank}(\boldsymbol{Z})=2 } \left[ \min_{\boldsymbol{e},\,\|\boldsymbol{e}\|_1= M \varsigma^2 }  \left\|\hat{\boldsymbol{X}} - \boldsymbol{Z}\right\|_2^2\right]\right],
\end{equation}
where $\hat{\boldsymbol{X}}$ denotes the reconstructed coordinates of nodes given the tempered EDM in \eqref{10} and \eqref{e}. In other words, the level is quantified by the minimum squared Euclidean distance from the reconstructed coordinate space to its projection into any $2$D space, given that the faulty node implements the attack that minimizes this distance. It is essential to note the sequence of minimization, meaning that the faulty node first chooses the error then the closest $2$D space is selected. Since this quantity is affected by the locations of good nodes by noting that closer good nodes produce stronger lever force in the seesaw test given the same strength of shout attack, the expectation in \eqref{test} is taken over a given location distribution. In the following theorem, we adopt the $2$D Gaussian distribution for ease of exposition. 
\begin{theorem}
\label{thm_seesaw}
Assume that the faulty node is at $(0,0)$ with the attack strength of $\|\boldsymbol{e}\|_1= M \varsigma^2$, and the good nodes' coordinates are i.i.d. generated based on a Gaussian distribution with zero mean and variance of $\sigma^2$, i.e., $(x_m,y_m) \sim \mathcal(\boldsymbol{0},\sigma^2 \boldsymbol{I})$, $\forall m \in {1,...,M}$. When the error $\boldsymbol{e}$ is independent with $(x_m,y_m)$, $\forall m \in {1,...,M}$, then 
\begin{equation}
\label{h}
    h (\varsigma^2) = \min\left\{(M-1)\sigma^2,\,(M-2)\varsigma^2\right\}.
\end{equation}
\end{theorem}
\begin{IEEEproof}
See Appendix \ref{app_seesaw}.
\end{IEEEproof}
\begin{remark}
It is shown that a faulty node cannot conceal its lie by noting that $h(\varsigma^2)$ scales with the attack strength $\varsigma^2$, until $\varsigma^2$ is comparable with the squared distance measurement ($\sigma^2$) whereby the attack becomes quite obvious. In addition, the effect is amplified by approximately $M$ times; this is intuitive since it becomes increasingly more difficult to lie to more good nodes when they form a concrete $2$D space. Another note is that $h(\varsigma^2)>0$ as long as $M \ge 3$, because at least $3$ nodes can determine a $2$D space.
\end{remark}
\begin{remark}
The theorem assumes that the error matrix $\boldsymbol{E}$ is independent with the coordinates of good nodes, which requires that a faulty node is not aware of the coordinates of other nodes (assumed mostly good nodes) in the sortition phase; this is reasonable because the coordinate information is not accessible in the sortition phase before any distance feedback occurs. 
\end{remark}
\begin{coro}
In $L$-dimensional space, $h (\varsigma^2) = \min\left\{(M-L+1)\sigma^2,\,(M-L)\varsigma^2\right\}$.
\end{coro}
\begin{remark}
This corollary generates the effectiveness of the seesaw test to higher dimensions, e.g., $3$D scenarios with applications for, e.g., drone swarms.
\end{remark}
\section{Byzantine Agreement}
\label{sec_ba}
In this phase, since we have removed Sybil nodes from the selected senators to a great extent, basically any byzantine agreement protocol can be implemented among the $K$ senators. In particular, we adopt the Jack scheme proposed in \cite{sto16} which consists of the following two major stages:
\begin{itemize}
    \item 
    \textbf{Setup stage}: Each senator broadcasts its initial value and receives other senators' initial values. Thereby, each senator broadcasts its acceptable values and a proposed value, jointly considering its and other senators' initial values.
    \item
    \textbf{Search stage}: Rotating among $(t+1)$ pre-determined leaders, in each round a leader receives proposals from other senators and accordingly proposes a value based on its acceptable values; if an agreement is reached based on proposals, the leader would propose the agreed value. It is proved that as long as one leader among the $(t+1)$ leaders is a good node, a valid agreement would be reached. Therefore, at most $t$ faulty nodes are allowed in this phase.
\end{itemize}

The validity is assured by the median validity specified in Section \ref{sec_model}. The termination and agreement properties are also proved in \cite{sto16}. In fact, this protocol achieves the optimum safety and $2$-approximate of the optimum median validity.

\section{Simulations}
\label{sec_sr}
We run a computer simulation to test the performance of SENATE. The ranging estimations at nodes are assumed to be perfect, and the information exchange is simulated by direct modifications of data arrays for nodes. 
\begin{figure}[!t]
	\centering
	\includegraphics[width=0.45\textwidth]{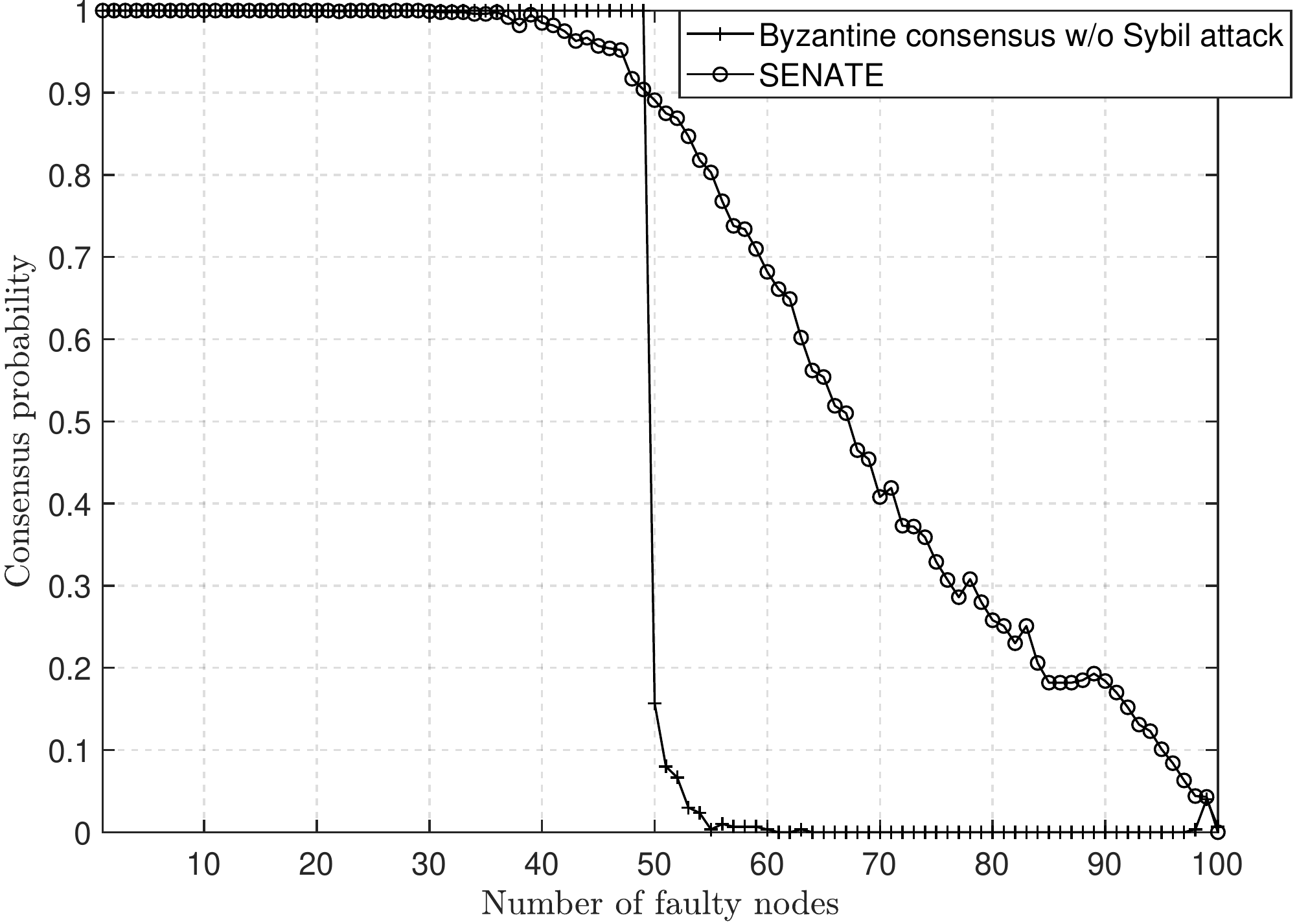}
	\caption{The probability a valid consensus is reached with $100$ nodes.}
	\label{Fig_prob}
\end{figure}

In Figure \ref{Fig_prob}, where there are $N=100$ nodes and SENATE selects $S=50$ candidates and $K=7$ senators, the number of faulty nodes is shown by x-axis and the performance of consensus probability is shown which is obtained by running the algorithms for $1000$ episodes. The node locations are randomly generated in a square with a side length of $200$ meter in each episode. The faulty nodes are assumed to always launch Sybil attacks and propose randomly generated values which are deviating from the true values; specifically, the good and faulty nodes' initial values are uniformly randomly generated from interval $[-1,1]$ and $[99,101]$, respectively. In the figure, we also simulate the Jack algorithm \cite{sto16} without faulty nodes launching Sybil attacks for comparisons; the Jack algorithm under this scenario and attack assumptions can ensure consensus when the number of faulty nodes does not reach majority ($50\%$), and cannot otherwise; this is direct consequences of the design of the Jack algorithm and the definition of median validity. It is observed that SENATE perform close-to the conventional BFT protocol as if there is no Sybil attack; this verifies that SENATE is Sybil-proof. We also observe that SENATE perform better when faulty nodes reach majority, which is because the SENATE randomly selects the senators such that there is a probability that the selected senators are dominated by good nodes; the Jack algorithm can also perform, at least, as well as SENATE if a sortition phase is added, but the effect is not shown in the figure to be in line with the original Jack algorithm. 

\section{Conclusion}
\label{sec_con}
SENATE is a real-time distributed BFT protocol which is applicable to fully-connected wireless-networked systems without prior identity authentication. In order to prevent malicious nodes to generate an arbitrary number of pseudonyms, SENATE leverages the wireless signals transmitted by nodes to cross-verify their identities in a fully-decentralized (no-trust) manner, based on the fact that pseudonyms are likely to be adjacent geographically. Thereby, only selected nodes, i.e., senators, participate in the final consensus reaching process. Computer simulations show that SENATE is Sybil-proof, by comparing the consensus probabilities between systems wherein faulty nodes launch and do not launch Sybil attacks, respectively. 

\section*{Acknowledgement}
This work is sponsored in part by the Nature Science Foundation of China (No. 61701275, No. 91638204, No. 61571265, No. 61621091), the China Postdoctoral Science Foundation, and Hitachi Ltd.

\appendices
\section{Proof of Theorem \ref{thm_nash}}
\label{app_nash}
First, we invoke the following lemma which ensures the existence of a Nash equilibrium.
\begin{lemma}[\cite{nash}]
\label{lm_nash}
A finite symmetric game has a symmetric mixed-strategy equilibrium.
\end{lemma}
The finite symmetric game in Lemma \ref{lm_nash} denotes a game wherein every player has the same finite action set, and the payoff received by each player given the same action (and other players' actions) is identical, irrespective with the specific player. A mixed strategy is in contrast with a pure strategy; the latter employs a fixed action each time whereas the former can be viewed as a mix (randomized strategy) of the latter. The proof of Lemma \ref{lm_nash} is based on the Brouwer fixed point theorem and is omitted for brevity.

Provided the existence of symmetric equilibria, we are ready to derive the transmission probability $p$ for every node. Due to symmetry, it suffices to consider a single node, whose game choice is shown in Table \ref{tab_game}.
\begin{table}[h!]
\label{tab_game}
\caption{Payoff functions}
\centering
\begin{tabular}{ |c|c|c|c| } 
 \hline
  & \tabincell{c}{All other nodes \\are silent $(p_1)$}   & \tabincell{c}{Some other node\\ transmits $(p_2)$} & Expected payoff\\ 
 \hline
 Transmit & $1-c$ & $-c$ & $(1-c)p_1 - cp_2$ \\ 
 \hline
 Silent & $0$ & $0$ & 0 \\ 
 \hline
\end{tabular}
\end{table}

The probability of each event is
\begin{IEEEeqnarray}{rCl}
p_1 = \Pr\{\mathsf{success}\}&=&(1-p)^{N-1}; \nonumber\\
p_2 = \Pr\{\mathsf{collision}\}&=&1-(1-p)^{N-1}.
\end{IEEEeqnarray}
Based on the principle of indifference \cite{Cheng04}, both expected payoffs should be zero, which yields \eqref{p_n} and concludes the proof.

\section{Proof of Theorem \ref{thm_seesaw}}
\label{app_seesaw}
Based on \eqref{9}, the EDM without error can be written as
\begin{IEEEeqnarray}{rCl}
    \boldsymbol{D} &=& -2\boldsymbol{X}\boldsymbol{X}^\mathsf{T} + \boldsymbol{1} \mathsf{diag} \left(\boldsymbol{X}\boldsymbol{X}^\mathsf{T}\right)^\mathsf{T} + \mathsf{diag} \left(\boldsymbol{X}\boldsymbol{X}^\mathsf{T}\right) \boldsymbol{1}^\mathsf{T} \nonumber\\
    &=& -2\boldsymbol{X}\boldsymbol{X}^\mathsf{T} + \boldsymbol{1}\left[0,\,\boldsymbol{\beta}_0^\mathsf{T}\right] + \left[0,\,\boldsymbol{\beta}_0^\mathsf{T}\right]^\mathsf{T}\boldsymbol{1}^\mathsf{T},
\end{IEEEeqnarray}
where $\boldsymbol{\beta}_0 \triangleq \left[d_{10}^2,...,d_{M0}^2\right]^\mathsf{T}$, and $\boldsymbol{X} = \left[ {\begin{array}{*{20}{c}}
0&{{x_1}}& \cdots &{{x_M}}\\
0&{{y_1}}& \cdots &{{y_M}}
\end{array}} \right]^\mathsf{T} \triangleq [\boldsymbol{0},\,\bar{\boldsymbol{X}}^\mathsf{T}]^\mathsf{T}$. The attack is implemented on EDM and we can derive the resultant coordinate covariance, which is tampered by the attack, based on \cite{dok15}:
\begin{IEEEeqnarray}{rCl}
\label{xxt}
&& \hat{\boldsymbol{X}} \hat{\boldsymbol{X}}^\mathsf{T} \nonumber\\
&=& -\frac{1}{2}\left(\boldsymbol{D} + \boldsymbol{E} - \boldsymbol{1}[0,\,\boldsymbol{\beta}_0^\mathsf{T} + \boldsymbol{e}^\mathsf{T}] - [0,\,\boldsymbol{\beta}_0^\mathsf{T}+\boldsymbol{e}^\mathsf{T}]^\mathsf{T}\boldsymbol{1}^\mathsf{T}\right) \nonumber\\
&=& {\boldsymbol{X}} {\boldsymbol{X}}^\mathsf{T} + \frac{1}{2}\left(
\left[ {\begin{array}{*{20}{c}}
0&\boldsymbol{0}^\mathsf{T}\\
\boldsymbol{0}&{\boldsymbol{1}\boldsymbol{e}^\mathsf{T}}
\end{array}} \right] + 
\left[ {\begin{array}{*{20}{c}}
0&\boldsymbol{0}^\mathsf{T}\\
\boldsymbol{0}&{\boldsymbol{e}\boldsymbol{1}^\mathsf{T}}
\end{array}} \right] \right) \nonumber\\
&=&  
\left[ {\begin{array}{*{20}{c}}
0&\boldsymbol{0}^\mathsf{T}\\
\boldsymbol{0}&{\bar{\boldsymbol{X}}\bar{\boldsymbol{X}}^\mathsf{T} + \frac{1}{2}\left(\boldsymbol{1}\boldsymbol{e}^\mathsf{T} + \boldsymbol{e}\boldsymbol{1}^\mathsf{T}\right)}
\end{array}} \right].
\end{IEEEeqnarray}
First, we notice that the eigenspace of $\hat{\boldsymbol{X}} \hat{\boldsymbol{X}}^\mathsf{T}$ in \eqref{xxt} is at most $4$-dimensional (the other eigenvalues are zeros) by the rank inequality
\begin{IEEEeqnarray}{rCl}
    \mathsf{rank}(\hat{\boldsymbol{X}} \hat{\boldsymbol{X}}^\mathsf{T}) &\le& \mathsf{rank}(\bar{\boldsymbol{X}}\bar{\boldsymbol{X}}^\mathsf{T})  + \mathsf{rank}(\boldsymbol{1}\boldsymbol{e}^\mathsf{T} )+ \mathsf{rank}(\boldsymbol{e}\boldsymbol{1}^\mathsf{T}) \nonumber\\
    &=& 4.
\end{IEEEeqnarray}
In addition, the eigenspace is spanned by 
\begin{equation}
\label{1e}
    \mathsf{eigenspace}(\hat{\boldsymbol{X}} \hat{\boldsymbol{X}}^\mathsf{T}) = \mathsf{span}\{\boldsymbol{x},\boldsymbol{y},\boldsymbol{1},\boldsymbol{e}\},
\end{equation}
where $\boldsymbol{x} \triangleq [x_1,...,x_K]^\mathsf{T}$ and $\boldsymbol{y} \triangleq [y_1,...,y_K]^\mathsf{T}$; this can be seen from the following equation.
\begin{IEEEeqnarray}{rCl}
&& \bar{\boldsymbol{X}}\bar{\boldsymbol{X}}^\mathsf{T} + \frac{1}{2}\left(\boldsymbol{1}\boldsymbol{e}^\mathsf{T} + \boldsymbol{e}\boldsymbol{1}^\mathsf{T}\right) \nonumber\\
&=& [\boldsymbol{x},\boldsymbol{y},\boldsymbol{1}, \boldsymbol{e}] \left[ {\begin{array}{*{20}{c}}
\boldsymbol{I}&\boldsymbol{0}\\
\boldsymbol{0}&{\frac{1}{2}\boldsymbol{I}}
\end{array}} \right] [\boldsymbol{x},\boldsymbol{y}, \boldsymbol{e},\boldsymbol{1}]^\mathsf{T}.
\end{IEEEeqnarray}
Note that we assume $\boldsymbol{E}$ is independent with $\bar{\boldsymbol{X}}$, and that the power of the error is $\varsigma^2$, i.e.,
\begin{equation}
    \left\|\frac{\boldsymbol{1}\boldsymbol{e}^\mathsf{T} + \boldsymbol{e}\boldsymbol{1}^\mathsf{T}}{2}\right\|_* =  \varsigma^2 M.
\end{equation}
Since the error is assumed to be independent with the coordinates and that i.i.d. Gaussian coordinate vectors are uniformly directed in space,  a constant share of the error power is leaked out of the coordinate eigenspace spanned by $[\boldsymbol{x},\boldsymbol{y}]$. Considering the objective of the faulty node is to minimize the leakage power beyond any $2$D space, which in this case is equivalent to maximizing the power in the $2$D space given that the total power is fixed, the best attack the faulty node can implement is to concentrate its error power to the same linear space spanned by $\boldsymbol{1}$, i.e., 
\begin{equation}
\label{eopt}
    \boldsymbol{e}_{\textrm{opt}} = \varsigma^2 \boldsymbol{1}.
\end{equation}
Note that the solution $\boldsymbol{e} = -\boldsymbol{e}_{\textrm{opt}}$ also satisfies the conditions, however, it results in a negative eigenvalue of $\hat{\boldsymbol{X}} \hat{\boldsymbol{X}}^\mathsf{T}$. This violates with \cite[Theorem 2]{gow82} which proves an important property of EDM that if $\hat{\boldsymbol{X}} \hat{\boldsymbol{X}}^\mathsf{T}$ is not positive-semi-definite, then there exists at least three distance measurements violating the triangle inequality. In other words, a shout attack may put the faulty node in a higher-dimensional space whereas a whisper attack would lead to violation of the triangle inequality which is much easier to spot.

Let us consider the minimization inside the expectation in \eqref{test}, this is an optimal low-rank approximation problem whose solution is well known to be the dominant $2$-dimensional singular space of $\hat{\boldsymbol{X}}$, i.e., the eigenspace of $\hat{\boldsymbol{X}} \hat{\boldsymbol{X}}^\mathsf{T}$. Denoting the singular value decomposition (SVD) of $\hat{\boldsymbol{X}}$ as $\hat{\boldsymbol{X}} = \hat{\boldsymbol{U}} \hat{\boldsymbol{\Sigma}} \hat{\boldsymbol{V}}^\mathsf{T}$ (singular values are always arranged in non-increasing order), then
\begin{equation}
    \boldsymbol{Z}_{\textrm{opt}} = \hat{\boldsymbol{U}}_2 \hat{\boldsymbol{\Sigma}}_2 \hat{\boldsymbol{V}}_2^\mathsf{T},
\end{equation}
wherein $\hat{\boldsymbol{U}}_2$ and $\hat{\boldsymbol{V}}_2$ denote the first two columns of $\hat{\boldsymbol{U}}$ and $\hat{\boldsymbol{U}}$, respectively, and $\hat{\boldsymbol{\Sigma}}_2$ contains the two dominant singular values. It follows that the minimum projection error is
\begin{IEEEeqnarray}{rCl}
\min_{\boldsymbol{Z} \in \mathbb{R}^{M \times M},\,\mathsf{rank}(\boldsymbol{Z})=2 } \left\|\hat{\boldsymbol{X}} - \boldsymbol{Z}\right\|_2^2 &=& \left\|\hat{\boldsymbol{X}} - \boldsymbol{Z}_{\textrm{opt}}\right\|_2^2 = \sum_{i=3}^M \Sigma_{i,i}^2, \nonumber\\
\end{IEEEeqnarray}
which is the coordinate power leakage beyond $2$D space due to tampered EDM. To solve for this quantity, we adopt the Gram–Schmidt orthogonalization process on the set $\{\boldsymbol{x},\boldsymbol{y}, |\varsigma|\boldsymbol{1}\}$ (given the optimal attack derived in \eqref{eopt}). Based on symmetry that $\boldsymbol{x}$ and $\boldsymbol{y}$ are both i.i.d. Gaussian distributed, it is clear that the direction of the third vector is not relevant and hence it is replaced by $\boldsymbol{w}$ where $\boldsymbol{w} \triangleq \sqrt{\varsigma^2M}[1,...,0]^{\mathsf{T}}$ with the same power. For brevity, the detailed process is omitted as the orthogonal basis vectors and the expected leakage power $h(\varsigma^2)$ are given as below:
\begin{IEEEeqnarray}{rCl}
 \boldsymbol{u}_1 &=& \boldsymbol{x}; \nonumber\\
 \boldsymbol{u}_2 &=& \boldsymbol{y} - \frac{\boldsymbol{u}_1^{\mathsf{T}}\boldsymbol{y} }{\|\boldsymbol{u}_1\|^2_2} \boldsymbol{u}_1; \nonumber\\
 \boldsymbol{u}_3 &=& \boldsymbol{w} - \frac{\boldsymbol{u}_1^{\mathsf{T}}\boldsymbol{w} }{\|\boldsymbol{u}_1\|^2_2} \boldsymbol{u}_1 - \frac{\boldsymbol{u}_2^{\mathsf{T}}\boldsymbol{w} }{\|\boldsymbol{u}_2\|^2_2} \boldsymbol{u}_2; \nonumber\\
 h(\varsigma^2) &=& \min_{i=\{1,2,3\}} \{\mathbb{E}[\boldsymbol{u}_i^{\mathsf{T}}\boldsymbol{u}_i]\} \nonumber\\
&=& \min_{i=\{1,2,3\}} \{M\sigma^2,(M-1)\sigma^2,(M-2)\varsigma^2\},
\end{IEEEeqnarray}
and the conclusion follows immediately.
\bibliography{senate}
\bibliographystyle{IEEEtran}

\end{document}